\documentclass[aps,titlepage,preprint]{revtex4-2}
\usepackage[utf8]{inputenc}
\usepackage[margin=1in]{geometry}
\usepackage{amsmath,bm}
\usepackage{xcolor}
\usepackage{braket}
\usepackage{mathrsfs}
\usepackage{amsfonts}
\usepackage{epsfig,float}
\usepackage[normalem]{ulem}
\usepackage{amssymb,amsfonts}
\usepackage{accents}

\definecolor{indigo(dye)}{rgb}{0.0, 0.25, 0.42}

\usepackage[colorlinks, allcolors=blue!70!black, linktocpage]{hyperref}
\usepackage{cleveref}
\crefname{section}{\S}{\S}
\crefname{appendix}{Appendix}{Appendices}
\crefname{figure}{Fig.}{Figs.}
\crefname{definition}{Def.}{Defs.}
\crefname{prop}{Prop.}{Props.}
\crefname{lemma}{Lemma}{Lemmas}
\crefname{corollary}{Cor.}{Cors.}
\crefname{thm}{Theorem}{Theorems}
\crefname{remark}{Remark}{Remarks}
\crefname{ass}{Assumptions}{Assumptions}
\crefname{property}{Properties}{Properties}

\makeatletter
\long\def\dddddot#1{%
  {\mathop {#1}\limits ^{\vbox to-1.4\ex@ {\kern -\tw@ \ex@ \hbox {\normalfont .....}\vss }}}%
}
\long\def\multidots#1#2{%
  \count@=0
  {{\mathop {#2}\limits ^{\vbox to-1.4\ex@ {\kern -\tw@ \ex@ \hbox {\normalfont %
  \loop%
  \ifnum#1>\count@%
  .%
  \advance\count@ by1%
  \repeat%
  }\vss }}}}%
}
\makeatother

\begin{document}

\title{On primordial gravitational waves in Teleparallel Gravity}
\author{Geovanny A. Rave-Franco}
\email{geovanny.rave@ciencias.unam.mx}
\affiliation{Instituto de Ciencias Nucleares, Universidad Nacional Aut\'{o}noma de M\'{e}xico, 
Circuito Exterior C.U., A.P. 70-543, M\'exico D.F. 04510, M\'{e}xico.}

\author{Celia Escamilla-Rivera}
\email{celia.escamilla@nucleares.unam.mx}
\affiliation{Instituto de Ciencias Nucleares, Universidad Nacional Aut\'{o}noma de M\'{e}xico, 
Circuito Exterior C.U., A.P. 70-543, M\'exico D.F. 04510, M\'{e}xico.}

%%%%%%%%%%%%%%%%%%%%%%%%%%%%%%%%%%%%%%%%%%%%
%%%%%%%%%%%%%%%%%%%%%%%%%%%%%%%%%%%%%%%%%%%%

\begin{abstract}
   Teleparallel Gravity is a gauge theory where gravity is a manifestation of the torsion of space-time and its success relies on being a possible solution to some problems of General Relativity. In this essay we introduce the construction of the theory by defining its geometrical setup, and how we can build it as a gauge theory of translations locally invariant under the Lorentz group. In this context, we will study the production of primordial gravitational waves and the observational implications when extended models are taken into account, particularly, we will notice how the tensor spectral index changes and produces a direct impact on the power spectrum from vacuum fluctuations and any source of tensor anisotropic stress in comparison to General Relativity.
   
   \vspace{24pt}
{\em Essay written for the Gravity Research Foundation 2023 Awards for Essays on Gravitation}

\end{abstract}

%%%%%%%%%%%%%%%%%%%%%%%%%%%%%%%%%%%%%%%%%%%%
%%%%%%%%%%%%%%%%%%%%%%%%%%%%%%%%%%%%%%%%%%%%

\maketitle

General Relativity is one of the most successful theories in physics, explaining gravitation in terms of the geometry of space-time and having numerous experimental confirmations, ranging from Solar system tests up to cosmological scales \cite{Clifton2012,Will2014}. Despite its success, this theory faces different issues, some of them are theoretical, like the existence of singularities \cite{Penrose1965}  or the quantization of gravity \cite{tHOOFT1993}, and others are observational like statistical tensions on cosmological models derived from it and over measurements of important cosmological parameters \cite{DiValentino2021}, the coincidence problem \cite{Velten2014}, among others. A possible and natural solution from first principles to some of these issues is the introduction of an extension of a theory of gravity so-called: \textit{Teleparallel Gravity}, a gauge theory of translations locally invariant under the Lorentz group where gravitation is due to torsion and not due to curvature like in General Relativity. Let us begin the discussion by introducing the geometrical setup of this theory, then proceed with a direct construction of the gauge structure of the theory followed by the construction of the gravitational action and its extensions, and finally discuss some of the modern studies done in the context of Teleparallel Gravity. \\

\noindent Teleparallel Gravity is constructed over the tangent bundle $TM$ given by
\begin{align}
TM = \left\lbrace (p,T_p \mathcal M) \in \mathcal M \times \bigcup_{p \in M}T_p \mathcal M: X_p \in T_p \mathcal M \right\rbrace,
\end{align}
where $ \mathcal M$ is the space time manifold and a fiber $T_p \mathcal M$ identified as the Minkowski space \cite{Aldrovandi2013}. Over both spaces local coordinates can be introduced, the use of Greek indices $\{x^{\mu}\}$ will refer to indices over the space-time manifold, and Capital Latin indices $\{x^A\}$ will refer to Minkowski indices. On this bundle, there are two fundamental geometrical objects, the tetrad fields $\boldsymbol{e} \in \Omega^1(\mathcal{M},\mathbb{R}^{1,3})$, $\boldsymbol{e} = \{e^A\}^{3}_{A=0}$, which are differential $1$-forms over $\mathcal{M}$ assuming values on the Minkowski space, and the spin connection $\boldsymbol{\omega} \in \Omega^1(\mathcal{M},\mathfrak{so}(1,3))$, a differential $1$-form over $\mathcal{M}$ assuming values on the Lie Algebra of the Lorentz group. The tetrad fields constitute a basis of the fields of differential $1$-forms, such that if $J$ is a differential $1$-form, then locally
\begin{align}
    e^A = e^{A}_{\ \mu}dx^{\mu}, \quad J = J_{\mu}dx^{\mu} \quad \longrightarrow \quad  J = J_{A}e^{A} = (J_Ae^{A}_{\ \mu})dx^{\mu}.
\end{align}
Defining the inverse of the tetrad fields as $E^{\mu}_{ \ A}$ such that $E^{\mu}_{\ A}e^{A}_{\ \nu} = \delta ^{\mu}_{\nu}$ and $E^{\mu}_{\ A}e^{B}_{\ \mu}= \delta ^{B}_{A}$, we have the transformation rule to obtain the coefficients of the $J$ $1$-form in terms of the new basis as
\begin{align}
    J_{A} = J_{\mu}E^{\mu}_{\ A}.
\end{align}
Analogously, the inverse of the tetrad fields constitutes a basis of the space of vector fields over $\mathcal{M}$, such that if $V$ is a vector field, locally it implies
\begin{align}
    E_{A} = E^{\mu}_{\ A}\partial_{\mu}, V = V^{\mu}\partial_{\mu} \quad \longrightarrow \quad V = V^{A}E_{A} = (V^AE^{\mu}_{\ A})\partial_{\mu},
\end{align}
from where the transformation rule can be obtained as $V^{A}=V^{ \mu}e^{A}_{\ \mu}$. In this new basis, the coefficients of the metric tensor will change as
\begin{align}
    g_{AB} = E^{\mu}_{\ A}E^{\nu}_{ \ B}g_{\mu \nu} \quad \text{or} \quad   g_{\mu \nu} = e^{A}_{\ \mu}e^{A}_{ \ \nu}g_{AB},
\end{align}
however, there are some particular tetrad fields called \textit{Orthonormal Fields}, these fields are the ones used in the Teleparallel Gravity literature since they require
\begin{align}
    g_{AB} \equiv \eta_{AB} \quad \longrightarrow \quad g_{\mu \nu} = e^{A}_{\ \mu}e^{A}_{ \ \nu}\eta_{AB},
\end{align}
allowing us to recover the Minkowskian space at each tangent space of the tangent bundle, as desired from the geometrical setup previously discussed. From now on, when talking of tetrads we will refer particularly to those orthonormal fields. From this definition, since the Minkowski metric is invariant under transformations of the Lorentz group $\text{SO}^+(1,3)$ and the space-time metric follows the usual transformation rule for tensors under diffeomorphisms, the tetrad field will follow the following transformation rules for Lorentz transformations and diffeomorphisms
\begin{align}
e^{A}_{\ \mu} = \Lambda ^{A}_{\ B}e ^{B}_{\ \mu} \quad \text{and} \quad  e^{A}_{\ \mu '} = \frac{\partial x^{\nu}}{\partial x^{\mu '}}e^{A}_{\ \nu } ,
\end{align}
respectively. On the other hand, the spin connection will summarize all the inertial effects of the theory, making the theory to be locally invariant under the Lorentz group. The spin connection is chosen such that the coefficients of the curvature $ 2$ form
\begin{align}\label{curvature 2-form}
\mathbf{R} = \frac{1}{4}R^A_{\text{   }B \nu \mu}S_A^B dx^{\nu}\wedge dx^{\mu}, \quad \text{with} \quad R^A_{\text{   }B\nu\mu} = \partial_{\nu}\omega^{A}_{\text{   }B\mu} - \partial_{\mu}\omega^{A}_{\text{   }B\nu} + \omega^A_{ \text{   }D \nu}\omega^D_{\text{   }B \mu} - \omega^A_{\text{   }D \mu}\omega^D_{\text{   }B \nu},
\end{align}
and $S^{AB}$ are the Lorentz generators for a given representation, are vanishing, this requires that the spin connection acquires a particular form called the \textit{purely inertial} spin connection
\begin{align}\label{purelyinertialspinconnection}
\omega^{A}_{\ B \mu} = \Lambda^{A}_{\ C}(x)\partial_{\mu}\Lambda^{C}_{\ B}(x),
\end{align}
for a particular local Lorentz transformation $\Lambda^{A}_{\ B}(x)$. This spin connection is the transformed zero spin connection given in a particular frame, under an arbitrary Lorentz transformation. In this scheme, Teleparallel Gravity requires that the torsion $2$-form is non-vanishing 
\begin{align}\label{torsion 2-form}
\mathbf{T} = \frac{1}{2}T^{A}_{\nu \mu}P_A dx^{\nu} \wedge dx^{\mu} \quad \text{with} \quad  T^A_{\text{   } \nu \mu} = \partial_{\nu} e^A_{\mu} - \partial_{\mu}e^A_{\nu} + \omega^A_{\text{   }C \nu}e^{C}_{\mu} - \omega^A_{\text{   }C \mu}e^C_{\nu} \neq 0,
\end{align}
with $P_A$ the generators of translations. The relation between the coefficients of the curvature and torsion $2$-forms with those of the curvature and torsion tensors are
\begin{align}\label{relations}
R^{\rho}_{\text{   }\gamma \nu \mu} = E^{\rho}_{\ A} e^{B}_{\ \gamma}R^A_{\text{   }B \nu \mu} \quad \text{and} \quad T^{\rho}_{\nu \mu}=E_{A}^{\rho}T^{A}_{\nu \mu} = -2\Gamma^{\rho}_{\left[ \nu \mu \right]},
\end{align}
with the linear connection given by 
\begin{align}\label{Connection}
\Gamma_{\nu \mu}^{\rho} \equiv E_{A}^{\ \rho}\partial_{\mu}e^{A}_{\ \nu} + e_{A}^{\rho}\omega^{A}_{\text{   }B \mu}e^{B}_{ \ \nu} = e_{A}^{\rho}\mathscr{D}_{\mu}e^{A}_{\ \nu}.
\end{align}Although this is the geometry of the theory, how we can construct the theory with such a gauge structure? This can be achieved by constructing a covariant derivative of scalar fields as follows: 
we begin by considering a local translation of the Minkowskian coordinates 
\begin{align}\label{traslations}
x^A \longrightarrow x'^A + \epsilon^A(x^{\mu}),
\end{align}
then, a Minkowskian scalar field will transform as
\begin{align}
\phi(x^A) \longrightarrow \phi(x'^A) = \phi(x^A - \epsilon^a)= \phi(x^A) - \epsilon^A\partial_{A}\phi,
\end{align}
from where its infinitesimal change is given by
\begin{align}\label{infinitesimal}
\delta \phi \equiv \phi(x^A) - \phi(x'^A) = \epsilon^A \partial_A \phi.
\end{align}
Analogously, the derivative $\partial_{\mu}\phi$ is also a Minkowskian scalar field, however, if we perform the local coordinate transformation given by Eq.~(\ref{traslations}), we will notice that 
\begin{align}
\delta(\partial_{\mu}\phi) = \epsilon^A \partial_A\partial_{\mu}\phi + (\partial_A\phi)(\partial_{\mu}\epsilon^A),
\end{align}
which does not follow the same rule found in Eq.~(\ref{infinitesimal}). This is where the introduction of a translation potential 
\begin{align}
B_{\mu} = B^A_{\mu} P_A,
\end{align}
is required, and such potential must satisfy $\delta B^A_{\mu} = -\partial_{\mu}\epsilon^A$, in order to allow the construction of the covariant derivative
\begin{align}
e_{\mu}\phi = \partial_{\mu}\phi + B^{A}_{\mu}\partial_A \phi,
\end{align}
that follows the correct transformation rule 
\begin{align}
\delta(e_{\mu}\phi) = \epsilon^A\partial_A\partial_{\mu}\phi.
\end{align}
Therefore, the gauge structure is introduced into the theory by replacing the regular partial derivative with the covariant derivative constructed above $\partial_{\mu} \to e_{\mu}$. We can identify the components of the tetrad field from the covariant derivative by identifying
\begin{align}
    e_{\mu} = e^{A}_{\ \mu}\partial_{A} \quad \longrightarrow \quad  e^{A}_{\ \mu} =  \partial_{\mu}x^A + B^A_{\mu}.
\end{align}
Finally, we require the theory to be invariant under Local Lorentz transformations, hence, if we perform the transformation
\begin{align}
x^A \longrightarrow \Lambda^A_{\ B}x^B \quad\text{and} \quad e^A_{ \ \mu} = \Lambda^A_{ \ B}e^B_{\ \mu},
\end{align}
it can be noticed that 
\begin{align}
\Lambda^A_{\ B} e^B_{\mu} = \partial_{\mu}(\Lambda^A_{\ B}x^B) + \Lambda^A_{\ C} B^C_{\mu} \quad \longrightarrow \quad  e^A_{\mu} =  \partial_{\mu}x^A + B^A_{\mu} + \omega^A_{\ B \mu}x^B,
\end{align}
where the purely inertial spin connection $ \omega^A_{\ B \mu} = \Lambda^A_{\ D}\partial_{\mu} \Lambda^D_{\ B}$, has explicitly appeared into the tetrad to guarantee the Lorentz invariance. \\
Thus, the theory has now been constructed to be a gauge theory of translation locally invariant under Lorentz transformations, and we are now in a position to try to study Cosmology within this theory. \\ 

At first, we can choose any tetrad-spin connection pair to determine the geometry of our tangent bundle, however in physics, particularly, in gravitational and cosmology topics, we need to satisfy the field equations for a particular scenario or symmetry of a problem. As a first step, we require to construct an action for this theory from where we can obtain field equations to solve. This can be easily achieved by considering the General Relativity action from the Einstein-Hilbert action \cite{Carroll2004}
\begin{equation}
\label{GR action}
S[g_{\mu \nu},\bm{\psi}]= \frac{1}{2\kappa}\int  \accentset{\circ}{R}\sqrt{-g} d^4x+ S_{\text{matt}}[g_{\mu \nu},\bm{\psi}],
\end{equation}
with $\kappa=8\pi G$, and use the definition of the contorsion tensor \cite{Bahamonde2023}
 \begin{equation}
\Gamma^{\rho}_{\mu \nu} = \accentset{\circ}{\Gamma} ^{\rho}_{\mu \nu} + K^{\rho}_{\mu \nu},
\end{equation}
with the over circle indicating that it is related to the Levi-Civita connection, i.e. that linear connection of General Relativity. From this, we can compute the Ricci scalar as
 \begin{align}\label{ricci torsion}
R = \accentset{\circ}{R} + T - B,
\end{align}
and since there is no curvature in Teleparallel Gravity, we can arrive at
\begin{align}\label{Ricci}
    \accentset{\circ}{R}= - T + B,
\end{align}
with the \textit{torsion scalar} $T$ and the \textit{boundary term} given by
\begin{align}
T &=T^{\alpha}_{\text{   }\sigma \rho}S_{\alpha}^{\text{   }\sigma \rho}= \frac{1}{4}T^{\mu \nu \lambda}T_{\mu \nu \lambda} + \frac{1}{2}T^{\mu \nu \lambda}T_{\nu \mu \lambda} - T^{\mu}T_{\mu},\label{torsionscalar} \\
S_{\alpha}^{\text{   }\sigma \rho} &= \frac{1}{4}(T_{\alpha}^{\text{   }\sigma \rho} + T^{\rho\sigma}_{\hspace{0.3cm} \alpha} - T^{\sigma \rho}_{\hspace*{0.3cm}\alpha} - 2 T^{\lambda\sigma}_{\hspace*{0.3cm}\lambda}\delta_{\alpha}^{\rho} + 2 T^{\lambda \rho}_{\hspace*{0.3cm}\lambda}\delta_{\alpha}^{\sigma}) = \frac{1}{2}\left( K_{\hspace{0.3cm} \alpha}^{\sigma \rho} +  T^{\sigma}\delta_{\alpha}^{\rho} -  T^{\rho}\delta_{\alpha}^{\sigma} \right), \label{superpotentialtensor} \\
B &= \frac{2}{e}\partial_{\mu}\left( e T^{\mu} \right) = 2\nabla_{\mu}T^{\mu}. \label{boundaryterm}
\end{align}
Using Eq.~(\ref{Ricci}) and the relation between the determinant of the tetrad and the metric $\sqrt{-g}=e$, we have 
\begin{align}
    S = \frac{1}{2\kappa}\int (-T + B)ed^4x + S_{\text{matter}},
\end{align}
but $B$ is a boundary term, satisfying $\int Bed^4x = 0$, then, we arrive at the so-called \textit{Teleparallel Equivalent to General Relativity} (TEGR)
\begin{align}
    S_{\text{TEGR}} = -\frac{1}{2\kappa} \int Ted^4x + S_{\text{matter}}.
\end{align}
From this action, we can compute the field equations associated with variations concerning the tetrad and the spin connection, which are completely equivalent to those of General Relativity. However, variations of the action concerning the spin connection are equal to the antisymmetric part of the field equations coming from variations of the action concerning the tetrad, and in the case of cosmological symmetry, the antisymmetric part of the field equations vanishes \cite{Hohmann2019} allowing us to only work with the tetrad as the fundamental field and work within the Weitzenböck gauge $\omega^{A}_{\ B \mu}=0$, this is known as the \textit{pure tetrad} formalism.  In this formalism, the field equations are given by 
\begin{align}\label{TEGR field equations}
W^{\lambda}_{\ \nu} \equiv \frac{2}{e}e^{A}_{\nu}\partial_{\mu}(e S_A^{\text{   }\lambda \mu}) - 2 T^{\alpha}_{\text{   }\mu \nu}S_{\alpha}^{\text{   }\mu \lambda} + \frac{1}{2}T \delta^{\lambda}_{\nu} =  \kappa \Theta^{\lambda}_{\ \nu},
\end{align}
which provide the same phenomenology as General Relativity, where the energy-momentum tensor is defined as
\begin{equation}\label{Energy-momentum tensor TG}
    \Theta^{\lambda}_{\ \nu} = e^{A}_{\ \nu}\theta^{\lambda}_{\ A}, \quad \theta_{\ A}^{\lambda} = \frac{1}{e} \frac{\delta (e \mathcal{L}_{\text{m}})}{\delta e^A_{\ \lambda}}\,.
\end{equation} This equivalence encourages the search for extensions of TEGR, since, as stated before, there are observational issues in GR that are also present in TEGR. Some of these extensions have been done in the form of general functionals of the torsion scalar, known as $f(T)$ gravity, or the torsion scalar and the boundary term, known as $f(T,B)$ gravity. The $f(T)$ gravity extension is just a limit of the $f(T,B)$ given by $f(T,B)=f(T)$, i.e. dropping out the dependence on the boundary term. This allows us to discuss the general case which is $f(T,B)$ knowing that the $f(T)$ case is only a subset of $f(T,B)$ gravity. The action for $f(T,B)$ gravity is given by 
\begin{align}\label{f(T,B) action}
S_{f(T,B)} = \frac{1}{2\kappa}\int ef(T,B)d^4x + S_{\text{matt}},
\end{align}
and the field equations are
\begin{align}\label{f(T,B) field equations}
    W^{\lambda}_{\ \nu} \equiv & \delta^{\lambda}_{\nu}\accentset{\circ}{\square}f_B -  \accentset{\circ}{\nabla}^{\lambda}\accentset{\circ}{\nabla}_{\nu}f_B + \frac{1}{2}f_BB\delta^{\lambda}_{\nu} + 2\left[\partial_{\mu}f_B + \partial_{\mu}f_T \right]S_{\nu}^{\text{   }\mu \lambda} + \frac{2}{e}e^A_{\nu}f_T\partial_{\mu}(e S_A^{\text{   }\mu \lambda}) \\ \notag & - 2f_TT^{\alpha}_{\text{   }\mu \nu}S_{\alpha}^{\text{   }\lambda \mu} - \frac{1}{2}f\delta^{\lambda}_{\nu} =  \kappa \Theta^{\lambda}_{\ \nu}\,.
\end{align}
Recently, a study in the context of the early universe using this theory will be reported in \cite{rave-etal}, particularly, the study of the production of primordial gravitational waves, an important prediction from inflation. In this cited work the observational implications of the presence of extended models appear in the power spectrum. The idea is to break the cosmological symmetry by introducing small deviations in the cosmological background 
\begin{align}
e^{A}_{\ \mu} = \bar e^{A}_{\ \mu} +  \delta e^{A}_{\ \mu},
\end{align}
at linear order, with the $(+,-,-,-)$ signature and the background cosmology in conformal time 
\begin{align}
    \bar e^{A}_{\ \mu} = a(\eta)\text{diag}(1,1,1,1).
\end{align}
Since the interest is to study primordial gravitational waves, the linear perturbation is a transverse and traceless tensor perturbation
\begin{align}\label{ptetradtensor}
\delta e^{A}_{\ \mu} = \frac{a(\eta)}{2}\begin{pmatrix}
0 & 0 \\
0 & h_{ij} \\
\end{pmatrix}, \quad \delta^{ij}h_{ij} = 0, \quad \partial^{i}h_{ij}=0.
\end{align}
The perturbed field equations, associated with this tensor perturbation, in Fourier space are
\begin{align}\label{GWTG}
h''_{kj} + [2 + \nu]\mathcal{H}h'_{kj}  + k^2 h_{kj} = \frac{16\pi G a^{2}}{f_T}\pi_{ij}^T,
\end{align}
where derivatives are with respect to the conformal time, $\nu = \frac{1}{\mathcal{H}}\frac{f'_T}{f_T}$ and $\pi_{ij}^T$ the tensor anisotropic stress. Two possible sources of these gravitational waves are considered, vacuum fluctuations and tensor anisotropies, and two backgrounds compatible with an inflationary era are considered, a perfect de Sitter background and a quasi de Sitter background. \\

The idea is to consider the beginning of the universe such that the gravitational wavefield is a quantum field \cite{Piattella2018}
\begin{eqnarray}\label{GWexpansionplanewaves}
	h_{ij}(\eta,\mathbf x) = \sum_{\lambda=\pm 2}\int\frac{d^3\mathbf k}{(2\pi)^3}\left[h(\eta,k)e^{i\mathbf k\cdot\mathbf x}a(\mathbf k,\lambda)e_{ij}(\hat{k},\lambda)+ h^*(\eta,k)e^{-i\mathbf k\cdot\mathbf x}a^{\dagger}(\mathbf k,\lambda)e^*_{ij}(\hat{k},\lambda)\right]\;, \quad
\end{eqnarray}
where the annihilation and creation operators satisfy the usual commutation rules
\begin{equation}\label{commutationrelationsGW}
	\left[a(\mathbf k,\lambda), a(\mathbf k',\lambda')\right] = 0\;, \qquad \left[a(\mathbf k,\lambda), a^\dagger(\mathbf k',\lambda')\right] = (2\pi)^3\delta^{(3)}(\mathbf k - \mathbf k')\delta_{\lambda\lambda'}\;.
\end{equation}
We now consider the quantum state of the universe to be a vacuum state 
\begin{align}
a(\boldsymbol{k},\lambda)\ket{0}=0,
\end{align}
whose initial states in the infinite past $\eta \to - \infty$ are those of a free field in Minkowksi space \cite{Piattella2018,Baumann2022-sw}. Such vacuum is called the \textit{Bunch-Davies vacuum} and the associated initial states are the Bunch-Davies mode functions. With this quantum state and the tensor field is given by eq. (\ref{GWexpansionplanewaves}), it is possible to compute the correlator of the field 
\begin{eqnarray}
	\langle 0|h_{ij}(\eta,\mathbf x)h_{lm}(\eta,\mathbf x')|0\rangle = \int\frac{d^3\mathbf k}{(2\pi)^3}|h(\eta,k)|^2e^{i\mathbf k\cdot(\mathbf x - \mathbf x')}\Pi_{ij,lm}(\hat{k})\;, \label{eq5}
\end{eqnarray}
where
\begin{equation}
	\Pi_{ij,lm}(\hat k) \equiv \sum_{\lambda = \pm 2}e_{ij}(\hat{k},\lambda)e^*_{lm}(\hat{k},\lambda)\;,
\end{equation}
and the power spectrum can be identified as
\begin{align}
    P_{h}(\eta,k)= |h(\eta,k)|^2.
\end{align}
It is possible then to study this power spectrum from different sources. Let us begin by considering vacuum quantum fluctuations. Vacuum fluctuations are present in the form of fluctuations in the inflating field, however, since the inflaton field is a scalar field its perturbations do not produce a tensor anisotropic stress. In the case of a perfect de Sitter background $H=$ cte, $\dot{H}=0$ the solution for the  $h(\eta,k)$ mode, is
\begin{align}
 h = \sqrt{32\pi G}\frac{1}{a}\frac{e^{-i k \eta}}{\sqrt{2k}}\left(1 - \frac{i}{k \eta} \right),
 \end{align}
 and then the dimensionless power spectrum turns out to be the same as General Relativity \cite{Boyle2008}
 \begin{align}
\Delta^2_h (\eta, k) \equiv \frac{d \bra{0}\hat{h}^2_{ik}\ket{0}}{d \ln k} = 64 \pi G \frac{k^3}{2\pi^2} |\hat h(\eta,k)|^2= \frac{2H^2_{\Lambda}}{\pi^2 M_{\text{pl}}^2}[1 + k^2\eta^2]  \quad  \accentset{k|\eta|\to 0}{\parbox{1cm}{\rightarrowfill}} \quad \frac{2 H_{\Lambda}^2}{\pi^2 M_{\text{pl}}^2 }, \label{DimensionlessGR}
 \end{align}
 with $\hat h$ the tensor field with canonical normalization $\hat h =  h / \sqrt{32\pi G}$ \cite{Klose2022b}. If the background is a quasi de Sitter expansion $\dot H = -\epsilon H_{\Lambda}^2$ with $\epsilon$ the first slow roll parameter, the solution of the Gravitational Wavefield is
 \begin{equation}
    h = \sqrt{32\pi G} a^{-(1+ \epsilon \gamma)}\frac{\sqrt{\pi}}{2}e^{i \frac{\alpha \pi }{2}  +  i\frac{\pi}{4}}\sqrt{|\eta|}H_{\alpha}^{(1)}(k|\eta|)\, ,
\end{equation}
and the dimensionless power spectrum is
\begin{equation}
    \Delta^2_h(\eta, k) = \frac{k^3 |\eta|}{\pi M_{\text{pl}}^2} a^{-2(1 + \epsilon \gamma)}|H_{\alpha}^{(1)}(k|\eta|)|^2\, \quad  \accentset{k|\eta|\to 0}{\parbox{1cm}{\rightarrowfill}} \quad \frac{2 H_{\Lambda}^2}{\pi^2 M_{\text{pl}}^2 }k^{-2\epsilon(1 + \gamma)}\,. \label{DimensionlessSpectrumTG}
\end{equation}
The power law dependence on the scale is called the \textit{tensor spectral index} $n_T = -2\epsilon(1+\gamma)$, with
\begin{align}
    \gamma = \left( \frac{f_{T_{\Lambda}T_{\Lambda}}|T_{\Lambda}| + f_{T_{\Lambda}B_{\Lambda}}|B_{\Lambda}|}{f_{T_{\Lambda}}}\right)\,.
\end{align}
This result is a key difference between General Relativity and Teleparallel Gravity extended models,  since extensions allow for a tensor spectral index with a high value, compared to the small value predicted by General Relativity given by $n_T= -2\epsilon$. If tensor anisotropies are included, the solution in a quasi-de Sitter background is 
\begin{eqnarray}
\label{GeneralSolutionTG}
    \hat h^{\lambda}_{>}(\eta,\boldsymbol{k}) &=& \int_{-\infty}^{\eta}d\eta_i \left[\frac{a(\eta_i)}{a(\eta)}\right]^{1 + \gamma \epsilon}G_{\text{TG}}^R(\eta, \eta_i,k) \hat \varrho_Q^{\lambda}(\eta_i,\boldsymbol{k})  
    \nonumber \\ &&
    - \int_{-\infty}^{\eta}d\eta_i \frac{1}{f_T(\eta_i)}\left[\frac{a(\eta_i)}{a(\eta)}\right]^{1 + \gamma \epsilon}G_{\text{TG}}^R(\eta, \eta_i,k) \varrho_T^{\lambda}(\eta_i,\boldsymbol{k})\,,
\end{eqnarray}
where the field has been split into a short wavelength mode plus a long wavelength mode $h(\eta,\boldsymbol{x}) = h_{<}(\eta,\boldsymbol{x}) + h_{>}(\eta,\boldsymbol{x})$, $G_{\text{TG}}^R(\eta, \eta_i,k)$ the Green's function of the gravitational waves propagation equation in Teleparallel Gravity in a vacuum, the quantum noise $\varrho_Q^{\lambda}$ and the tensor anisotropic noise $\varrho_T^{\lambda}$ are given by
\begin{align}
    \hat \varrho_{Q} = - \left({\hat h}''_{<} - \frac{2}{\eta}\left[1 + \epsilon(1+ \gamma) \right]{\hat h}'_{<} - \nabla^2 \hat h_{<} \right)\,, \quad  \varrho_T^{\lambda} = -16\pi G a^{2}\sum_{ij}\epsilon_{ij}^{\lambda *}\pi_{ij}^T.
\end{align}
From the general solution given in Eq.~(\ref{GeneralSolutionTG}) it is possible to obtain the power spectrum but the result is complicated and obtaining observational implications can be cumbersome, however, it is possible to analyze the observational implication in the $\epsilon \to 0$ limit as we shall see. From the general solution the power spectrum acquires the form $\Delta^2_h(\eta,k) = \Delta^2_{\text{vacuum}}(\eta,k) + \Delta^2_T(\eta,k)$ where $\Delta^2_{\text{vacuum}}(\eta,k)$ contribution is the same found in eq. (\ref{DimensionlessSpectrumTG}) and the contribution from tensor anisotropies $\Delta^2_T(\eta,k)=(k^3/2\pi^2)P_T(\eta,k)$ in the $\epsilon \to 0$ limit is
\begin{align}
     \delta_{D}(\boldsymbol k - \boldsymbol q)P_{T}(\eta,k)=&\frac{1}{f_T^2}\int_{-\infty}^{\eta} d\eta_{i}G_{R}^2(\eta,\eta_i,k)\sum_{\lambda}\bra{0} \varrho_T^{\lambda}(\eta_i,\boldsymbol{k}) \varrho_T^{\lambda}(\eta_i,\boldsymbol{q})\ket{0} \equiv \frac{1}{f_T^2}  \delta_{D}(\boldsymbol k - \boldsymbol q)P_T^{GR}(\eta,k)\,,
\end{align} 
with $P_T^{GR}(\eta,k)$ the power spectrum coming from tensor anisotropies in the context of General Relativity and $G_{R}(\eta,\eta_i,k)$ the Green's function of General Relativity. Several cosmological viable extended models satisfy $f_T < -1$ \cite{Bahamonde2023}, which implies that the peaks of the power spectrum coming from tensor anisotropies in Teleparallel Gravity will be suppressed by a factor of $1/f_T^2$ compared with the peaks in General Relativity. Some other studies in this line of thought involve stochastic gravitational wave background at second order \cite{Papanikolaou:2022hkg}.
Therefore, \textit{Teleparallel Gravity is a theory that can be used to explore the beginning of the universe with directly and significant observational implications}, in this case, on the primordial gravitational waves background that could be observed with future gravitational wave experiments like \href{https://lisa.nasa.gov/}{LISA} or the \href{https://www.et-gw.eu}{Einstein Telescope}. \\

\textit{Acknowledgments.-} 
GRF acknowledges financial support from SEP–CONACYT postgraduate grants program.
CE-R acknowledges the Royal Astronomical Society as FRAS 10147 and by DGAPA-PAPIIT-UNAM Project TA100122. This article is based upon work from COST Action CA21136 Addressing observational tensions in cosmology with systematics and fundamental physics (CosmoVerse) supported by COST (European Cooperation in Science and Technology).

\bibliographystyle{utphys}

\bibliography{ref}

\providecommand{\href}[2]{#2}\begingroup\raggedright\begin{thebibliography}{10}

\bibitem{Clifton2012}
T.~Clifton, P.~G. Ferreira, A.~Padilla, and C.~Skordis, ``Modified gravity and
  cosmology,'' \href{http://dx.doi.org/10.1016/j.physrep.2012.01.001}{{\em
  Physics Reports} {\bf 513} (2012) no.~1-3, 1--189}.
  \url{https://doi.org/10.1016/j.physrep.2012.01.001}.

\bibitem{Will2014}
C.~M. Will, ``The confrontation between general relativity and experiment,''
  \href{http://dx.doi.org/10.12942/lrr-2014-4}{{\em Living Reviews in
  Relativity} {\bf 17} (2014) no.~1, }.
  \url{https://doi.org/10.12942/lrr-2014-4}.

\bibitem{Penrose1965}
R.~Penrose, ``Gravitational collapse and space-time singularities,''
  \href{http://dx.doi.org/10.1103/physrevlett.14.57}{{\em Physical Review
  Letters} {\bf 14} (1965) no.~3, 57--59}.
  \url{https://doi.org/10.1103/physrevlett.14.57}.

\bibitem{tHOOFT1993}
G.~'t~HOOFT and M.~VELTMAN,
  \href{http://dx.doi.org/10.1142/9789814539395_0001}{``One-loop divergencies
  in the theory of gravitation,''} in {\em Euclidean Quantum Gravity},
  pp.~3--28.
\newblock {WORLD} {SCIENTIFIC}, May, 1993.
\newblock \url{https://doi.org/10.1142/9789814539395_0001}.

\bibitem{DiValentino2021}
E.~D. Valentino {\em et al.}, ``Snowmass2021 - letter of interest cosmology
  intertwined {II}: The hubble constant tension,''
  \href{http://dx.doi.org/10.1016/j.astropartphys.2021.102605}{{\em
  Astroparticle Physics} {\bf 131} (2021)  102605}.
  \url{https://doi.org/10.1016/j.astropartphys.2021.102605}.

\bibitem{Velten2014}
H.~E.~S. Velten, R.~F. vom Marttens, and W.~Zimdahl, ``Aspects of the
  cosmological {\textquotedblleft}coincidence problem{\textquotedblright},''
  \href{http://dx.doi.org/10.1140/epjc/s10052-014-3160-4}{{\em The European
  Physical Journal C} {\bf 74} (2014) no.~11, }.
  \url{https://doi.org/10.1140/epjc/s10052-014-3160-4}.

\bibitem{Aldrovandi2013}
R.~Aldrovandi and J.~G. Pereira,
  \href{http://dx.doi.org/10.1007/978-94-007-5143-9}{{\em Teleparallel
  Gravity}}.
\newblock Springer Netherlands, 2013.

\bibitem{Carroll2004}
S.~Carroll, {\em Spacetime and Geometry: An Introduction to General
  Relativity}.
\newblock Addison Wesley, 2004.

\bibitem{Bahamonde2023}
S.~Bahamonde {\em et al.}, ``Teleparallel gravity: from theory to cosmology,''
  \href{http://dx.doi.org/10.1088/1361-6633/ac9cef}{{\em Reports on Progress in
  Physics} {\bf 86} (2023) no.~2, 026901}.
  \url{https://doi.org/10.1088/1361-6633/ac9cef}.

\bibitem{Hohmann2019}
M.~Hohmann, L.~J\"{a}rv, M.~Kr{\v{s}}{\v{s}}{\'{a}}k, and C.~Pfeifer,
  ``Modified teleparallel theories of gravity in symmetric spacetimes,''
  \href{http://dx.doi.org/10.1103/physrevd.100.084002}{{\em Physical Review D}
  {\bf 100} (2019) no.~8, }. \url{https://doi.org/10.1103/physrevd.100.084002}.

\bibitem{rave-etal}
G.~A. Rave-Franco, C.~Escamilla-Rivera, and J.~L. Said, ``Production of
  primordial gravitational waves in teleparallel gravity,'' {\em Soon to be
  released} (2023)  .

\bibitem{Piattella2018}
O.~Piattella, \href{http://dx.doi.org/10.1007/978-3-319-95570-4}{{\em Lecture
  Notes in Cosmology}}.
\newblock Springer International Publishing, 2018.
\newblock \url{https://doi.org/10.1007/978-3-319-95570-4}.

\bibitem{Baumann2022-sw}
D.~Baumann, {\em Cosmology}.
\newblock Cambridge University Press, Cambridge, England, June, 2022.

\bibitem{Boyle2008}
L.~A. Boyle and P.~J. Steinhardt, ``Probing the early universe with
  inflationary gravitational waves,''
  \href{http://dx.doi.org/10.1103/physrevd.77.063504}{{\em Physical Review D}
  {\bf 77} (2008) no.~6, }. \url{https://doi.org/10.1103/physrevd.77.063504}.

\bibitem{Klose2022b}
P.~Klose, M.~Laine, and S.~Procacci, ``Gravitational wave background from
  non-abelian reheating after axion-like inflation,''
  \href{http://dx.doi.org/10.1088/1475-7516/2022/05/021}{{\em Journal of
  Cosmology and Astroparticle Physics} {\bf 2022} (2022) no.~05, 021}.
  \url{https://doi.org/10.1088/1475-7516/2022/05/021}.

\bibitem{Papanikolaou:2022hkg}
T.~Papanikolaou, C.~Tzerefos, S.~Basilakos, and E.~N. Saridakis, ``{No
  constraints for f(T) gravity from gravitational waves induced from primordial
  black hole fluctuations},''
  \href{http://dx.doi.org/10.1140/epjc/s10052-022-11157-4}{{\em Eur. Phys. J.
  C} {\bf 83} (2023) no.~1, 31}, \href{http://arxiv.org/abs/2205.06094}{{\tt
  arXiv:2205.06094 [gr-qc]}}.

\end{thebibliography}\endgroup
\end{document}